\documentclass[10pt,final,journal,letter,oneside,twocolumn]{IEEEtran}
\usepackage{cite}
\usepackage{ulem}
\usepackage{soul,color}
\usepackage{srcltx}
\usepackage{multirow}
\usepackage[tbtags,sumlimits,nointlimits,reqno]{amsmath}
\usepackage{graphicx,dblfloatfix}
\usepackage{float}
\usepackage{subfigure}
\usepackage{psfrag}
\usepackage{color}
\usepackage{amssymb}
\usepackage{makecell}
\usepackage{array}
\usepackage{epstopdf}
\usepackage{mathtools}
\usepackage{multirow}
\usepackage{threeparttable}

\begin{document}

\title{Low-Profile Spoof Surface Plasmon Polaritons Traveling-Wave Antenna for Endfire Radiation}

\author{Abhishek~Kandwal,~\IEEEmembership{Member,~IEEE},
        Qingfeng~Zhang,~\IEEEmembership{Senior Member,~IEEE},
        Xiao-Lan~Tang,\\
        Louis WY.~Liu,
        Zhang~Ge,~\IEEEmembership{Student~Member,~IEEE}

\thanks{Corresponding author: Qingfeng Zhang (email: zhang.qf@sustc.edu.cn).}
\thanks{The authors are with the Department of Electronics and Electrical Engineering, Southern University of Science and Technology of China, Shenzhen, Guangdong, China 518055. }

}

\maketitle

\begin{abstract}

This paper proposes a low-profile and highly efficient endfire radiating travelling-wave antenna based on spoof surface plasmon polaritons (SSPPs) transmission line. The aperture is approximately $0.32\lambda_0\times0.01\lambda_0$ where $\lambda_0$ is the space wavelength at the operational frequency 8 GHz. This antenna provides an endfire radiation beam within 7.5-8.5 GHz. The maximum gain and total efficiency reaches 9.2 dBi and $96\%$, respectively. In addition to the endfire operation, it also provides a beam scanning functionality within 9-12 GHz. Measurement results are finally given to validate the proposed SSPPs antenna.

\end{abstract}

\IEEEoverridecommandlockouts


\IEEEpeerreviewmaketitle

\section{Introduction}

Surface plasmons (SPs) are coherent electron oscillations at a metal-dielectric interface at optical frequencies~\cite{SP}. When SPs are coupled with EM waves (photons), the hybrid excitation is called surface plasmon polaritons (SPPs). So SPPs have both the charge oscillations in the metal and the evanescent EM waves in the dielectric. Although SPs and SPPs were discovered at optical frequencies, they have close connections with the surface waves at radio and microwave frequencies, i.e. Zenneck wave~\cite{zenneck1907propagation}, Sommerfeld wave~\cite{sommerfeld1909propagation} and Goubau line~\cite{goubau1950surface}.
In 2004, Pendry \emph{et al.} demonstrated a similar surface plasmonic phenomenon, known as spoof surface plasmon polaritons (SSPPs), at microwave frequencies~\cite{spoof_SSP}. Since then, SSPPs have attracted great attentions in the microwave community.

Owing to the groundless structure and strong field confinement in a subwavelength scale, SSPPs are good candidates for high-density integrated circuits and components at millimeter-wave and Terahertz frequencies~\cite{Ref20}. Extensive explorations and researches have been conducted in this area, including transition structures using coplanar waveguide (CPW) or microstrip line~\cite{2Liao2014, 3Kianinejad2016,4Yuan2016,5Kianinejad2015}, filters\cite{6Zhao2016,7Qian2016,8Xu2016}, amplifiers~\cite{Ref9}, power splitters~\cite{9Jin2014}, and antennas~\cite{11Yin2015,17Yin2016,10Zheng2016,13Xu2015,14Gu2016, 15Kianinejad2016,16Kianinejad2017,Ref21}.

SSPPs support transversal magnetic (TM) surface waves, which are slow-wave bounded modes that do not radiate due to the phase mismatch with the free space. In order to excite a radiation into the free space, one either uses tapering structures (mimicking horn antennas), or loads resonating elements (mimicking resonant antennas) or employs periodic modulations (mimicking non-resonant leaky-wave antennas). In~\cite{11Yin2015}, a flaring metal structure is used for impedance matching with the free space to achieve a wideband antenna. In~\cite{17Yin2016,10Zheng2016}, SSPPs loaded with circular patch arrays are reported for high gain and wideband operations. In~\cite{13Xu2015}, SSPPs with periodic profile modulations, inspired by phase-gradient metasurface, are proposed to design travelling-wave leaky-wave antennas. These leaky-wave antennas are capable of beam scanning in the whole space, but suffer from an efficiency degradation at the broadside. To overcome this problem, asymmetrical profile modulations are proposed for continuous scanning through broadside~\cite{14Gu2016}. A single-layer SSPPs meader line is also proposed for achieving a consistant gain in the beam scanning~\cite{16Kianinejad2017}. In addition to the leaky-wave antennas applications, SSPPs are also used to excite the fundamental horizontally polarized mode of dielectric resonator antennas~\cite{15Kianinejad2016}.

All the SSPPs leaky-wave antennas reported to date only steer the beam around the broadside, and very few studies have been dedicated to endfire radiations except the one in~\cite{Ref21} where SSPPs-fed dipole-like antenna was designed for endfire radiations. However, this endfire antenna exhibit a narrow frequency band due to the resonant nature of the dipole-like radiator. In this paper, we propose a travelling-wave endfire antenna using SSPPs structure. Due to the travelling-wave nature, it exhibit a wide operational frequency band. The subwavelength field confinement and single layer feature of SSPPs also lead to the low-profile nature of this antenna. Furthermore, it exhibits both endfire radiation and beam scanning capability, enabling it to be a multi-functional antenna.

\section{Principle}

\begin{figure}[!t]
  \centering
  \psfrag{w}[c][c]{\footnotesize $w$}
  \psfrag{T}[c][c]{\footnotesize $T$}
  \psfrag{L}[c][c]{\footnotesize $L$}
  \psfrag{g}[c][c]{\footnotesize $g$}
  \psfrag{d}[r][c]{\footnotesize $d$}
  \psfrag{p}[c][c]{\footnotesize $p$}
  \psfrag{h}[j][c]{\footnotesize $h$}
  \psfrag{z}[c][c]{\footnotesize $z$}
  \psfrag{x}[l][c]{\footnotesize $x$}
  \psfrag{y}[c][c]{\footnotesize $y$}
  \psfrag{a}[l][c]{\footnotesize $a$}
  \psfrag{1}[r][c]{\footnotesize \textcolor[rgb]{0.50,0.25,0.00}{Transition}}
  \psfrag{2}[c][c]{\footnotesize \textcolor[rgb]{0.50,0.25,0.00}{Guiding Area}}
  \psfrag{3}[c][c]{\footnotesize \textcolor[rgb]{0.50,0.25,0.00}{Tapering Area}}
  \includegraphics[width=8.6cm]{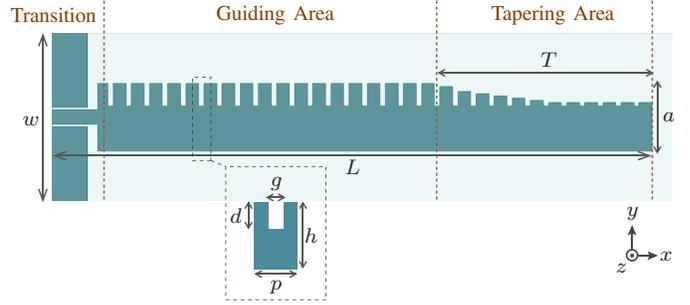}\\
  \caption{Configuration of the proposed SSPPs endfire antenna.}\label{fig:antenna_config}
\end{figure}

The configuration of the proposed SSPPs endfire antenna is shown in Fig.~\ref{fig:antenna_config}. It is composed of three main parts including transition, guiding area and tapering area. A transition from coplanar waveguide (CPW) to SSPPs is empolyed for mode conversion and impedance matching. The SSPPs transmission line in the guiding area supports a bounded surface-wave mode, which radiates at the end of a tapering SSPPs transmission line. The tapering SSPPs line provides a better impedance matching with the free space and hence significantly suppresses the reflection from the end. The antenna is printed on a Rogers Duroid 5880 substrate ($\epsilon_r=2.2, \tan\delta=0.0009$) with thickness $t=0.5$ mm. The dimensions are $w=30$ mm, $L=107$ mm, $a=12$ mm, $g=1$ mm, $h=12$ mm, $p=6$ mm, $d=4$ mm, and $T=42$ mm. The effective aperture of this endfire antenna is $a\times t=12~\text{mm}\times0.5~\text{mm}$, approximately $0.32\lambda_0\times0.01\lambda_0$ where $\lambda_0$ is the space wavelength at the operational frequency 8 GHz.

\begin{figure}[!t]
  \centering
  \psfrag{a}[l][c]{\footnotesize $d$ = 2 mm}
  \psfrag{b}[l][c]{\footnotesize $d$ = 3 mm}
  \psfrag{c}[l][c]{\footnotesize $d$ = 4 mm}
  \psfrag{d}[l][c]{\footnotesize $d$ = 5 mm}
  \psfrag{e}[l][c]{\footnotesize Air}
  \psfrag{x}[c][c]{\footnotesize Phase (degree)}
  \psfrag{y}[c][c]{\footnotesize Frequency (GHz)}
  \psfrag{f}[l][c]{\footnotesize $d$}
  \psfrag{g}[c][c]{\footnotesize PBC}
  \includegraphics[width=7cm]{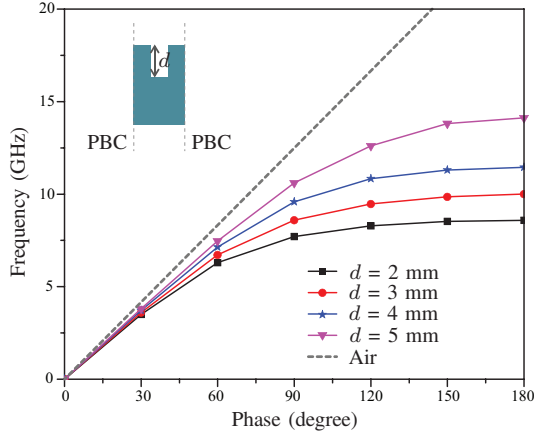}\\
  \caption{Dispersion curve of the unit cell in the guiding part of the antenna in Fig.~\ref{fig:antenna_config} ($g=1$ mm, $h=12$ mm, $p=6$ mm).}\label{fig:dispersion}
\end{figure}

The traveling-wave endfire antennas have two design guidelines~\cite{book_zucker}. Firstly, the dominant mode should be a bounded slow-wave mode so that the guided wave does not leak out until the end. Secondly, the phase velocity should be close to the light speed in the air so that the guided wave can smoothly radiate out at the end. Equivalently, the dispersion curve should stay below the air line within a short distance. Fig.~\ref{fig:dispersion} shows the dispersion curve of the unit cell in the guiding part of the proposed antenna in Fig.~\ref{fig:antenna_config}. It is analysed by enforcing periodic boundary conditions (PBC) on both sides of the unit cell and is computed using Eigenmode Solver of CST Microwave Studio. Note that, the dominant mode is a bounded slow-wave mode and the dispersion curve can be tuned by changing the groove depth $d$. In our case ($d=4$~mm), the dispersion curve around $8$ GHz is closely below the air line, which satisfies the design guidelines in~\cite{book_zucker}. The parameter $d=4$~mm is an optimum dimension for the maximum radiation gain at 8 GHz, which will be illustrated in the forthcoming paragraph.

\begin{figure}[!t]
  \centering
  \psfrag{A}[l][c]{\footnotesize A}
  \psfrag{1}[l][c]{\footnotesize A'}
    \psfrag{B}[l][c]{\footnotesize B}
  \psfrag{2}[l][c]{\footnotesize B'}
    \psfrag{C}[l][c]{\footnotesize C}
  \psfrag{3}[l][c]{\footnotesize C'}
  \psfrag{a}[c][c]{\footnotesize A-A'}
  \psfrag{b}[c][c]{\footnotesize B-B'}
  \psfrag{c}[c][c]{\footnotesize C-C'}
  \includegraphics[width=8cm]{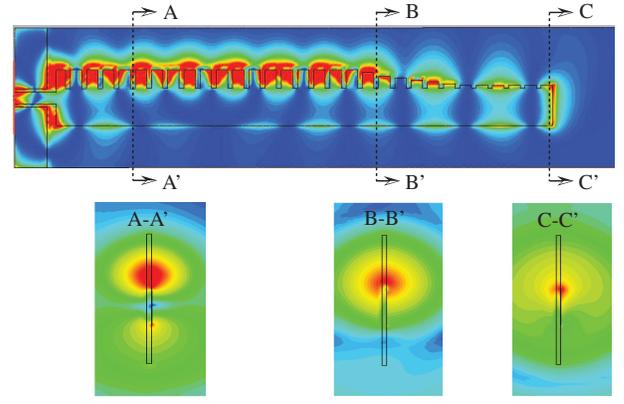}\\
  \caption{Electric field distribution at 8 GHz.}\label{fig:field_distribution}
\end{figure}

\begin{figure}[!t]
  \centering
    \psfrag{z}[c][c]{\footnotesize $z$}
  \psfrag{x}[c][c]{\footnotesize $x$}
  \psfrag{y}[c][c]{\footnotesize $y$}
  \includegraphics[width=8cm]{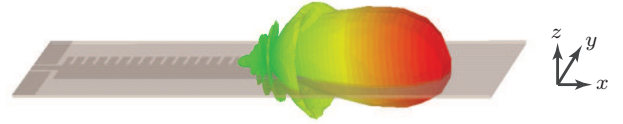}\\
  \caption{3-D radiation pattern of the SSPPs antenna at 8 GHz (in linear scale).}\label{fig:3d_pattern}
\end{figure}

\begin{figure}[!t]
  \centering
  \psfrag{x}[c][c]{\footnotesize $d$ (mm)}
  \psfrag{y}[c][c]{\footnotesize Peak Gain (dBi)}
  \psfrag{u}[c][c]{\footnotesize $L/\lambda_0$}
  \psfrag{v}[c][c]{\footnotesize Peak Gain (dBi)}
  \psfrag{a}[c][c]{\footnotesize (a)}
  \psfrag{b}[c][c]{\footnotesize (b)}
  \includegraphics[width=9cm]{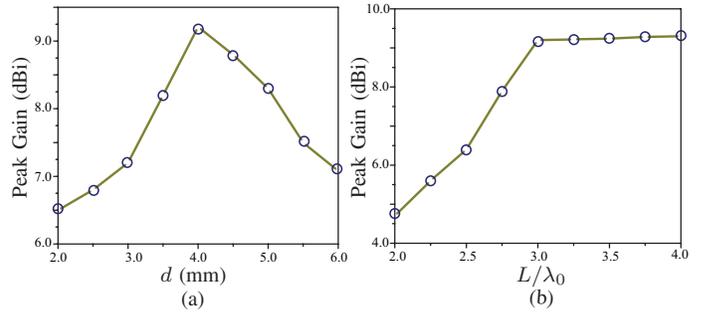}\\
  \caption{Peak gain versus (a) different groove depth $d$ (b) and overall length $L/\lambda_0$ ($g=1$ mm, $h=12$ mm, $p=6$ mm).}\label{fig:gain_optim}
\end{figure}

Fig.~\ref{fig:field_distribution} shows the electric field distribution along the antenna and at different cross-sections. From the distributions it is observed that the field is well confined and guided by the SSPPs line, satisfying the endfire radiaiton requirement. It is also observed that the field at the end is similar to that of a dipole and the effect of the tapered end is similar to the directing parasitic conductors in a yagi dipole antenna.
The 3D radiation pattern for the proposed SSPPs travelling-wave endfire antenna is shown in Fig.~\ref{fig:3d_pattern}. It can be clearly seen from this pattern that the antenna generates an endfire radiation beam at 8 GHz. Fig.~\ref{fig:gain_optim} shows the endfire gains for different groove depths and overall lengths. As $d$ gradually increases, the peak gain increases until $d=4$~mm (corresponding to a peak gain of 9.2 dBi) and then decreases. One may explain this using the two design guidelines in~\cite{book_zucker}. On one hand, as $d$ increases, the dispersion curve moves toward the air line, as shown in Fig.~\ref{fig:dispersion}, leading to a better phase matching with the free space for radiation at the end and hence an increasing gain. On the other hand, as $d$ further increases after some point, the field confinement capability of the SSPPs transmission line gets poor, leading to a decreasing gain. Therefore, there is an optimum groove depth (in our case $d=4$~mm) once the operational frequency and other dimensions are fixed. Note from Fig.~\ref{fig:gain_optim}(b) that, as the overall $L$ increases, the peak gain increases and then saturates after $L=3\lambda_0$ where $\lambda_0$ is the space wavelength at 8 GHz. Therefore, the optimum antenna length should be around $3\lambda_0$.

\begin{figure}[!t]
  \centering
  \psfrag{x}[c][c]{\footnotesize Frequency (GHz)}
  \psfrag{y}[c][c]{\footnotesize Reflection (dB)}
  \psfrag{a}[l][c]{\footnotesize Without Taper}
  \psfrag{b}[l][c]{\footnotesize With Taper}
  \includegraphics[width=7cm]{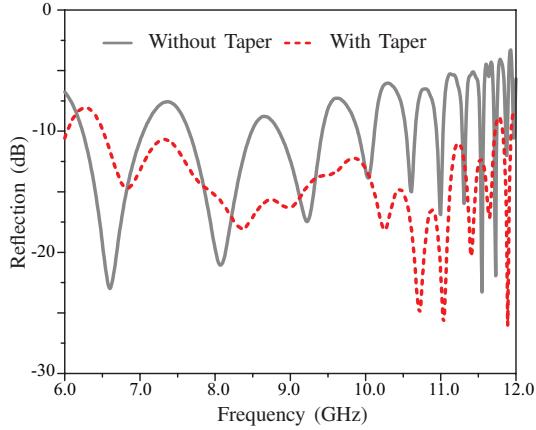}\\
  \caption{Reflections comparison of the endfire SSPPs antenna with and without the tapering end.}\label{fig:taper}
\end{figure}

The tapering end in the antenna generally offers a better impedance matching with the free space and hence suppresses the reflection. Fig.~\ref{fig:taper} compares the reflections of the antennas with and without the tapering end. Note that, with the tapering end, the reflections are suppressed below $-10$~dB within a wide frequency band from $6.5$~GHz to $11.5$~GHz. In contrast, the reflection without the tapering end is relatively poor.

\section{Experimental validation}

\begin{figure}[!t]
  \centering
  \psfrag{a}[l][c]{\footnotesize $g=1, d=4, h=12, p=2.5, a=12, w=30$}
  \psfrag{b}[l][c]{\footnotesize $L=107, T=42$}
  \psfrag{c}[r][c]{\footnotesize Unit: mm}
  \includegraphics[width=8cm]{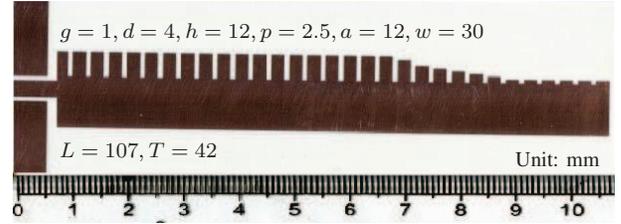}\\
  \caption{Fabricated prototype of the proposed SSPPs endfire antenna.}\label{fig:prototype}
\end{figure}

\begin{figure}[!t]
  \centering
  \psfrag{x}[c][c]{\footnotesize Frequency (GHz)}
  \psfrag{y}[c][c]{\footnotesize Reflection (dB)}
  \psfrag{a}[l][c]{\footnotesize Simulated}
  \psfrag{b}[l][c]{\footnotesize Measured}
  \includegraphics[width=7cm]{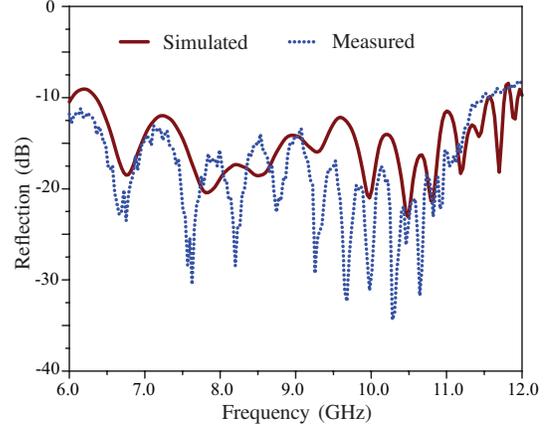}\\
  \caption{Measured and simulated reflection responses of the SSPPs antenna.}\label{fig:s-parameter}
\end{figure}

\begin{figure*}[!t]
  \centering
  \psfrag{a}[c][c]{\footnotesize (a)}
  \psfrag{b}[c][c]{\footnotesize (b)}
  \psfrag{c}[c][c]{\footnotesize (c)}
  \psfrag{d}[c][c]{\footnotesize $xy$-Plane}
  \psfrag{e}[c][c]{\footnotesize $yz$-Plane}
  \includegraphics[width=16cm]{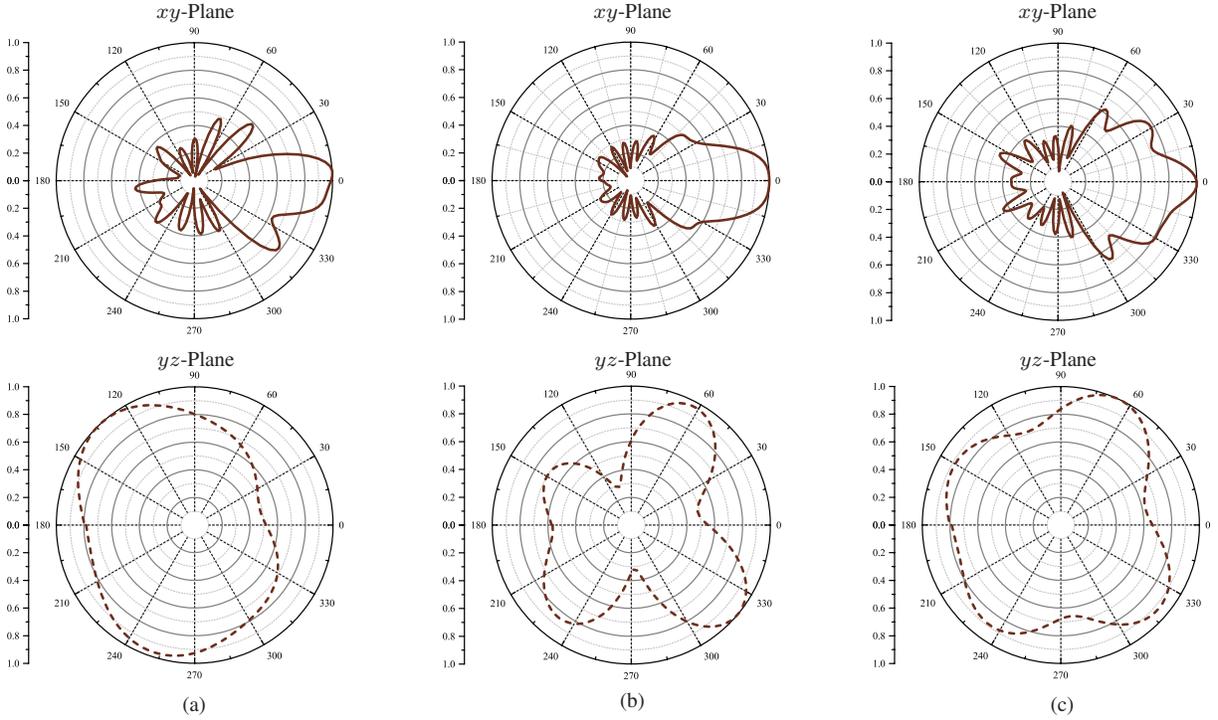}\\
  \caption{Measured radiation patterns ($xy$-plane and $yz$-plane) of the Endfire SSPPs antenna at (a) 7.5 GHz, (b) 8 GHz and (c) 8.5 GHz, respectively.}\label{fig:rad}
\end{figure*}

\begin{figure}[!t]
  \centering
  \psfrag{x}[c][c]{\footnotesize Frequency (GHz)}
  \psfrag{y}[c][c]{\footnotesize Peak gain (dBi)}
  \psfrag{z}[c][c]{\footnotesize Efficiency (\%)}
  \psfrag{a}[l][c]{\footnotesize Gain}
  \psfrag{b}[l][c]{\footnotesize Efficiency}
  \includegraphics[width=7cm]{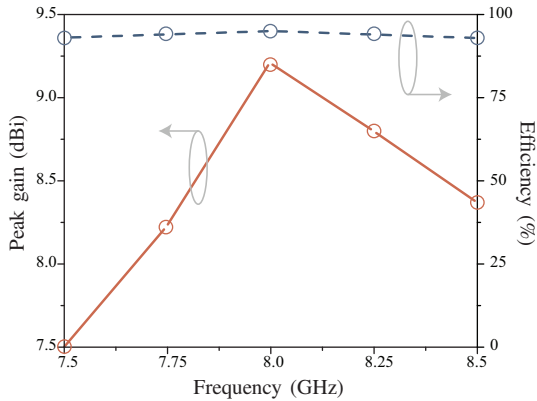}\\
  \caption{Peak gains and total efficiencies of the proposed antenna within 7.5-8.5 GHz.}\label{fig:fig10}
\end{figure}

To experimentally verify the proposed SSPPs endfire antenna in Fig.~\ref{fig:antenna_config}, we fabricate the prototype, as shown in Fig.~\ref{fig:prototype}. The reflection response is measured in an Agilent vector network analyzer E5051A, and the radiation pattern is measured using Satimo Starlab. Fig.~\ref{fig:s-parameter} shows the measured and simulated reflection responses, which agree well with each other. The measured reflection is suppressed below $-10$ dB within 6-11.5 GHz, which is even better than the simulated one.
Fig.~\ref{fig:rad} shows the measured radiation patterns at 7.5 GHz, 8 GHz, and 8.5 GHz, respectively. Note that, they are all endfire radiations at $xy$-plane and almost omnidirectional radiations at $yz$-plane. The proposed SSPPs antenna operates as an endfire antenna within the frequency range from 7.5 GHz to 8.5 GHz. Fig.~\ref{fig:fig10} plots the peak gains and the total efficiencies within 7.5-8.5 GHz. The total gain varies from 7.5 dBi to 9.2 dBi, with a maximum value of 9.2 dBi at 8 GHz. The total efficiency is almost constant, around $96\%$, within the whole endfire frequency range.

\begin{figure}[!t]
  \centering
  \psfrag{a}[l][c]{\footnotesize 9 GHz}
  \psfrag{b}[l][c]{\footnotesize 10 GHz}
  \psfrag{c}[l][c]{\footnotesize 12 GHz}
  \includegraphics[width=5.5cm]{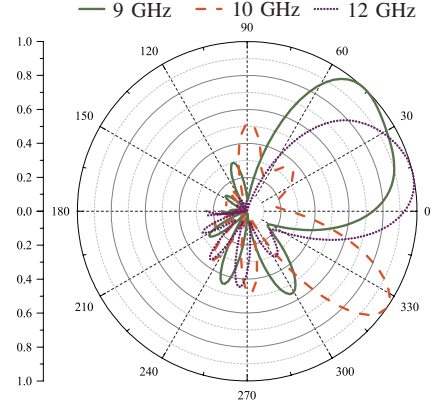}\\
  \caption{Scanning beams ($xy$-plane) of the proposed antenna at 9 GHz, 10 GHz and 12 GHz, respectively.}\label{fig:fig11}
\end{figure}

In addition to the endfire radiation within 7.5-8.5 GHz, the proposed SSPPs antenna also exhibits beam scanning functionality at other frequencies. Fig.~\ref{fig:fig11} shows the scanning radiation patterns at 9 GHz, 10 GHz and 12 GHz, respectively. Note that, the radiation beam scanns between $-35^{\circ}$ and $50^{\circ}$ around the endfire direction ($0^{\circ}$) within the frequency range from 9 GHz to 12 GHz. This is possibly attributed to the leaky-mode excited at higher frequencies. In summary, the proposed SSPPs antenna can operate as a multi-functional antenna at different frequencies.

\section{Conclusions}

 This paper has presented a low-profile and efficient SSPPs endfire antenna. This antenna has been successfully used to generated an endfire radiation within 7.5-8.5 GHz. A high efficiency of about 96 percent and a peak gain of 9.2 dBi has been obtained with this antenna. Ease of fabrication, low profile, minimized aperture and high efficiency rendered this antenna highly suitable for endfire applications.

\bibliographystyle{IEEEtran}
\bibliography{mybib}

\end{document}